\documentclass[superscriptaddress,preprint]{revtex4}
\usepackage{mathrsfs}
\usepackage{mathrsfs}
\usepackage{mathrsfs}
\usepackage{mathrsfs}
\usepackage{amssymb}
\usepackage[tbtags]{amsmath}
\usepackage{graphicx}
\usepackage{epsfig,graphicx,times}
\usepackage{color}
\usepackage{subfigure}

\begin{document}
\title{Quantum gates implementations in the separated ion-traps by fast laser pulses}
\author{ZHANG Miao}
\affiliation{Quantum Optoelectronics Laboratory, School of Physics
and Technology, Southwest Jiaotong University, Chengdu 610031,
China}
\author{WEI Lian-Fu\footnote{weilianf@mail.sysu.edu.cn; weilianfu@gmail.com}} \affiliation{Quantum Optoelectronics Laboratory, School of
Physics and Technology, Southwest Jiaotong University, Chengdu
610031, China} \affiliation{State Key Laboratory of Optoelectronic
Materials and Technologies, School of Physics and Engineering, Sun
Yat-sen University, Guangzhou 510275, China}\affiliation{State Key
Laboratory of Functional Materials for Informatics, Shanghai
Institute of Microsystem and Information Technology, Chinese Academy
of Sciences, Shanghai 200050, China}
\date{\today}


\begin{abstract}
An approach is proposed to implement the universal quantum gates
between the ions confined individually in the separated traps.
Instead of the typical adiabatic operations, performed for
manipulating the ion-ion coupling, here the switchable couplings
between ions are implemented non-adiabatically by using the fast
laser pulses. Consequently, the desirable quantum gates between the
ions could be implemented by using only a series of laser pulses.
The proposal may be conveniently generalized to the quantum
computation with the scalable ion-traps.

PACS numbers: 03.67.Lx, 42.50.Dv, 37.10.Ty
\end{abstract}

\maketitle

Among the proposed physical systems for building the future quantum
computer~\cite{QC}, the system of trapped
ions~\cite{Cirac,warmION,Physreports} is currently one of the most
hopeful candidates. Indeed, the entanglement up to fourteen trapped
ions has been experimentally realized~\cite{fourteenION}. In the
usual implementations a string of ions are trapped in a single trap,
and their center-of-mass mode (CM) acts as the data bus to couple
the qubits encoded by the long-lived electronic
levels~\cite{Physreports}.
However, it should be a challenge to trap more ions in a single well
and then coherently manipulate them precisely, as it is difficult to
obtain the reasonable strong-radial-confinements to form a linear
ion-string~\cite{Physreports}. Furthermore, the speed of the
laser-induced sideband excitation should be significantly reduced
with the larger mass of ion-chain.

Alternatively, the scheme of the segmented ion-traps could be an
more practical way to implement the large-scale quantum computing
network~\cite{scalable1,scalable2,Cirac2000}. This is because that
the ions can be relatively-easily confined in the separated
potential wells, and their couplings can be realized by transporting
the ions to any selected interaction regions. Specially, if the
separate wells are sufficiently close, then the trapped ions can be
coupled directly via their Coulomb interaction, and the ion
transporting is avoided. Interestingly, such an idea has been
demonstrated in recent experiments~\cite{experiment1,experiment2},
wherein two ions are confined in two potential wells separated by
$40\,\mu$m~\cite{experiment1} (or $54\,\mu$m~\cite{experiment2}) and
the ion-ion coupling is achieved up to $g\approx10$~kHz (or
$g\approx7$~kHz).

In a recent work~\cite{Miao} we have proposed an approach to
generate the universal quantum gates between the two separately
trapped ions. That idea is based on the
experiments~\cite{experiment1,experiment2}: the coupling between the
ions was manipulated by controlling the potential wells (i.e., the
voltages on the DC electrodes) to adiabatically tune ions' external
vibrations into or out of resonance. However, the adiabatic
conditions limit the speed of the quantum gates implementing. Also,
the exactly-controlling of the potential wells, by adjusting the
voltages applied on the DC electrodes, should be very complex for
the large-scale ion-traps.

In this letter we propose an improved approach to implement the
universal quantum gates between the separately trapped ions. Instead
of the previous adiabatic operations applied on the potential
wells~\cite{experiment1,experiment2,Miao}, here we use the laser
beams to drive one of the trapped ions for controlling its
interaction with another ion, let them be resonance or
large-detuning. With the present switchable couplings the desirable
coherent operations, e.g., the quantum gates, between the separately
trapped ions could be performed conveniently by using a series of
laser pulses.

\begin{figure}[tbp]
\includegraphics[width=9cm]{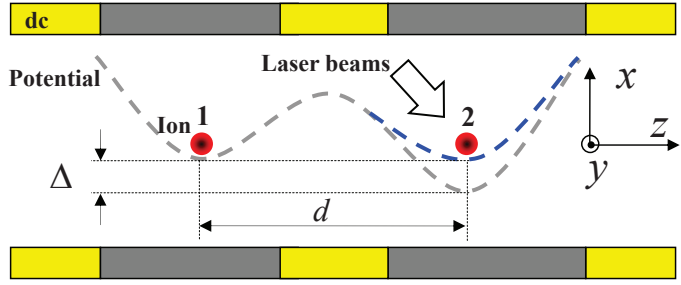}
\caption{(Color online) Sketch for implementing switchable coupling
between two ions trapped individually in two potential wells
(separated by the distance $d$). The ions are trapped initially in
the large-detuning regime, i.e., they are decoupled at the
beginning, then a laser pulse (applied to the ion $2$) makes
$\Delta=0$ for generating a resonant coupling.}
\end{figure}

We consider the same experimental configuration (shown in Fig.~1) as
that in the previous works~\cite{experiment1,experiment2,Miao},
i.e., two ions trapped individually in two potential wells separated
by a distance $d$. The ion trap provides a pseudopotential such that
the ions' oscillating frequencies along the axial direction (i.e.,
$z$ axis in the Fig.~1) is much smaller than those along the radial
directions (i.e., the $x,\,y$ axis). As a consequence, only the
quantized vibrational motion along the axial direction is
considered, and the Hamiltonian describing the oscillating ions
reads~\cite{experiment1,experiment2,Miao}
\begin{equation}
\hat{H}_{\rm
ex}=\sum_{j=1}^2\hbar\nu_j(\hat{a}_j^\dagger\hat{a}_j+\frac{1}{2})
+\hat{V}_{\rm ii}
\end{equation}
with
\begin{equation}
\hat{V}_{\rm ii}=K\left[\sum_{j=1}^2\frac{(-1)^{j}\xi_j}{d}\hat{a}_j
+\frac{\xi_j^2}{d^2}(\hat{a}_j^2+\hat{a}_j\hat{a}_j^{\dagger})
-\frac{2\xi_1\xi_2}{d^2}
(\hat{a}_1\hat{a}_2+\hat{a}_1\hat{a}_2^\dagger)\right] +\rm {H.c.}
\end{equation}
being the ion-ion Coulomb interaction. Where, $\hat{a}_j^\dagger$
and $\hat{a}_j$ are the bosonic creation and annihilation operators
of the vibration of frequency $\nu_j$. The parameters $K$ and
$\xi_j$ are defined as $K=q_1q_2/(4\pi\epsilon_0d)$ and
$\xi_j=\sqrt{\hbar/(2M_j\nu_j)}$, with $M_j$ and $q_j$ being the
mass and charge of the ion, respectively.

For realizing a switchable coupling between the two trapped ions,
without changing the eigenfrequencies $v_j$ of the ions, we apply
the laser beams to one of the ions, e.g., the ion $2$. The
Hamiltonian describing the laser-driving ion takes the well-known
form~\cite{IONSD}
\begin{equation}
\hat{H}_{\rm L}=\frac{\hbar\omega_a}{2}\hat{\sigma}_z+
\hbar\Omega_0(\hat{\sigma}_++\hat{\sigma}_-)
\left[e^{i\eta(\hat{a}_2+\hat{a}_2^\dagger)-i\omega_lt-i\phi_l} +\rm
{H.c.}\right].
\end{equation}
Here, $\hat{\sigma}_z=|e\rangle\langle e|-|g\rangle\langle g|$,
$\hat{\sigma}_+=|e\rangle\langle g|$, and
$\hat{\sigma}_-=|g\rangle\langle e|$ are the Pauli operators of the
ion's two internal states $|g\rangle$ and $|e\rangle$ (with the
transition frequency $\omega_a$ between them); $\Omega_0$ is the
Rabi frequency describing the strength of the coupling between the
applied lasers and the ion; $\eta$ is the Lamb-Dicke (LD) parameter
describing the strength of the coupling between the external and
internal states of the ion; and $\omega_l$ and $\phi_l$ are the
effective frequency and initial phase of the applied laser fields,
respectively. If the LD parameter $\eta$ is sufficiently small, then
the Hamiltonian in Eq. (3) can be approximated to
\begin{equation}
\hat{H}_{\rm L}'=\frac{\hbar\omega_a}{2}\hat{\sigma}_z+
\hbar\Omega_0(\hat{\sigma}_++\hat{\sigma}_-)
\left[(1+i\eta(\hat{a}_2+\hat{a}_2^\dagger)
-\frac{\eta^2}{2}(\hat{a}_2+\hat{a}_2^\dagger)^2)
e^{-i\omega_lt-i\phi_l}+\rm {H.c.}\right]
\end{equation}
by neglected the higher-order terms $O(\eta^3)$~\cite{SC}.

Consequently, the total Hamiltonian of the present two-ion system
$\hat{H}= \hat{H}_{\rm ex}+\hat{H}'_{\rm L}$
can be effectively approximated as
\begin{equation}
\hat{H}'=\frac{\hbar\delta_{\rm {in}}}{2}\hat{\sigma}_z
+\hbar\tilde{\Omega}\hat{F}+ \hbar(\delta_{\rm {ex}}-\Omega\hat{F})
\hat{a}_2^\dagger\hat{a}_2 -\hbar
g(\hat{a}_1\hat{a}_2^\dagger+\hat{a}_1^\dagger\hat{a}_2),
\end{equation}
in the rotating framework defined by the unitary operator
$\hat{U}_1=\exp[-it\omega_l\sigma_z/2
-it\sum_{j=1}^2(\nu_1\hat{a}_j^\dagger\hat{a}_j+\nu_j/2)]$. Above,
$\hat{F}=\exp(-i\phi_l)\hat{\sigma}_++\exp(i\phi_l)\hat{\sigma}_-$,
$\delta_{\rm ex}=\nu_2-\nu_1$, $\delta_{\rm in}=\omega_a-\omega_l$,
$\tilde{\Omega}=1-\eta^2/2$, $\Omega=\Omega_0\eta^2$, and
$g=2K\xi_1\xi_2/d^2$.
Suppose further that the ion is driven resonantly, i.e.,
$\delta_{\rm {in}}=0$, then the Hamiltonian in Eq. (5) reduces to
\begin{equation}
\hat{H}''=\hbar(\delta_{\rm {ex}}-\Omega\hat{F})
\hat{a}_2^\dagger\hat{a}_2 -\hbar
g(\hat{a}_1\hat{a}_2^\dagger+\hat{a}_1^\dagger\hat{a}_2)
\end{equation}
in another rotating framework defined by the unitary operator
$\hat{U}_2=\exp(-it\tilde{\Omega}\hat{F})$. Specially, if the
effective phase of the laser beams is set as $\phi_l=0$ and the
internal state of the ion is prepared in the eigenstate
$|\varphi\rangle=(|g\rangle+|e\rangle)/\sqrt{2}$ of the operator
$\hat{F}$, then the Hamiltonian in Eq. (6) reduces to
\begin{equation}
\hat{H}_{\rm {eff}}=\hbar\Delta\hat{a}_2^\dagger\hat{a}_2 -\hbar
g(\hat{a}_1\hat{a}_2^\dagger+\hat{a}_1^\dagger\hat{a}_2)
\end{equation}
with $\Delta=\delta_{\rm ex}-\Omega$ being the laser-induced
controllable detuning between the vibrations. In the previous
works~\cite{experiment1,experiment2,Miao} switchable couplings
between the ions were implemented by adiabatically manipulating the
vibrational frequencies $\nu_j$ (i.e., $\delta_{\rm ex}$). Here, we
fixed $\nu_j$ (and thus $\delta_{\rm ex}$ is unchanged) but just
adjust the laser-induced effective Rabi frequency $\Omega$ to
manipulate $\Delta$. Since $\Omega$ is relatively-easily controlled
by the applied laser and the adiabatic condition is unnecessary to
be satisfied, the laser-induced ion-ion coupling/decoupling may be
relatively-fast implemented.

\begin{figure}[tbp]
\includegraphics[width=9cm]{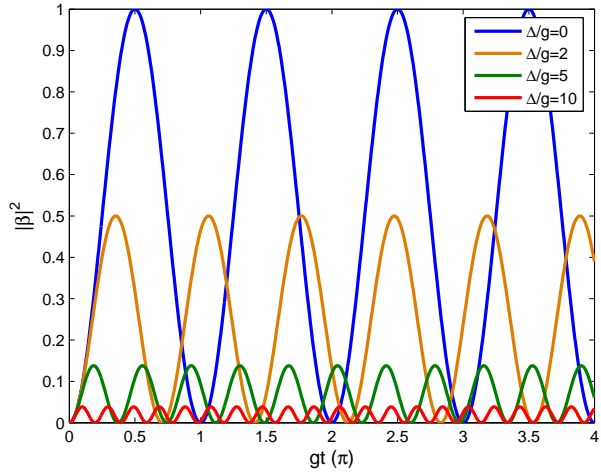}
\caption{(Color online) Occupancies $|\beta|^2$ in Eq. (8) versus
$gt$ for different detunings: $\Delta/g=0$, $\Delta/g=2$,
$\Delta/g=5$, and $\Delta/g=10$. The results indicate that the
probability, of the SWAP between the vibrational quanta of the two
ions, enhances with the decrease of detuning $\Delta$.}
\end{figure}

The effective Hamiltonian in Eq. (7) yields the dynamics for the
bosonic modes
\begin{eqnarray}
\begin{array}{l}
|0\rangle_1|1\rangle_2\longrightarrow \alpha|0\rangle_1|1\rangle_2
+\beta|1\rangle_1|0\rangle_2,\\
|1\rangle_1|0\rangle_2\longrightarrow \alpha|1\rangle_1|0\rangle_2
+\beta|0\rangle_1|1\rangle_2
\end{array}
\end{eqnarray}
with
\begin{equation}
\alpha=\left(\frac{1}{2}-\frac{\Delta}{4c}\right)
e^{i(c-\Delta/2)t}+
\left(\frac{1}{2}+\frac{\Delta}{4c}\right)e^{-i(c+\Delta/2)t},
\end{equation}
\begin{equation}
\beta=\frac{g}{c}e^{i(\pi-\Delta t)/2}\sin(ct),
\end{equation}
and $c=\sqrt{(\Delta/2)^2+g^2}$. Above, only two lowest-energy
vibrational states of each ion are considered and thus the dynamics
is limited within the invariant subspace: $\{|0\rangle_1|1\rangle_2,
|1\rangle_1|0\rangle_2\}$, with $|0\rangle_j$ and $|1\rangle_j$
being the ground- and first-excitation states of the $j$th ion's
vibration. The complex parameters $\alpha$ and $\beta$ satisfy the
normalized condition $|\alpha|^2+|\beta|^2=1$, where $|\beta|^2$ is
just the probability of the SWAP between the vibrational quanta of
the ions. Fig. 2 shows how the occupancy $|\beta|^2$ changes with
$gt$ for different $\Delta$, i.e., $\Delta=0, 2g, 5g$, and $10g$,
respectively. When $t\sqrt{(\Delta/2)^2+g^2}=n\pi/2$ with $ n=0, 1,
2,...$, the value of $|\beta|^2$ reaches its maximum (amplitude),
the maximum probability of the desirable SWAP. It is seen that such
an amplitude enhances with the decrease of detuning $\Delta$.
Therefore, one can turn on or off the coupling between the ionic
vibrations by setting $\Delta\approx0$ or $\Delta\gg g$,
correspondingly.
Specifically, such a switchable coupling can be implemented as:
Initially, the two ions are trapped in separate wells with the
detuning $\Delta=\delta_{\rm ex}\gg g$, i.e., their vibrations are
decoupled (shown, e.g., in Fig.~1). Then, a laser pulse with the
effective Rabi frequency $\Omega=\delta_{\rm ex}$ is applied to one
of the two ions for realizing the resonant coupling (i.e.,
$\Delta=0$). The induced evolution reads
\begin{equation}
\hat{U}_{1,2}(gt)=\left(
  \begin{array}{cc}
    0 & \cos(gt)\\
    i\sin(gt) & 0\\
  \end{array}
\right).
\end{equation}
After this laser driving the vibrations of the ions return to their
initial decoupled case.

In the above analysis we have assumed $g/\gamma\gg 1$ and neglected
the dissipation (with the rate $\gamma$) from the environment. Based
on the experiments~\cite{experiment1,experiment2}, we consider that
two $^{40}\rm {Ca}^+$ ions are trapped in different potential wells
(separated by the distance $d=40$~$\mu$m), and their vibrational
frequencies are set as $\nu_1=5$~MHz and $\nu_2=5.1$~MHz. In this
case, $g\approx11$~kHz and $\delta_{\rm ex}=\nu_2-\nu_1=0.1$~MHz,
which means that the two ions are effectively decoupled from each
other. Also, the typical decoherent rate of the ionic vibration is
$\gamma_{\rm ex}\sim1$~kHz~\cite{experiment1,experiment2}, and the
selected electronic levels, $|g\rangle=|\rm {S}_{1/2}\rangle$ and
$|e\rangle=|\rm {D}_{5/2}\rangle$, are long-lived with the
decoherent rate $\gamma_{\rm
in}\sim1$~Hz~\cite{quadrupole,quadrupoleCPL}. Therefore, the
condition $g/\gamma\gg1$ required in our proposal is satisfied. We
consider the quadrupole transition (at $729$~nm) between $|\rm
{S}_{1/2}\rangle$ and $|\rm {D}_{5/2}\rangle$~\cite{CPL729}.
Consequently, the LD parameter is calculated as $\eta\approx0.1$,
working within the LD regime. Experimentally, a laser beam of power
$P=140$~mW can generate a quadrupole transition between $|\rm
{S}_{1/2}\rangle$ and $|\rm {D}_{5/2}\rangle$ with Rabi frequency
$\Omega_0=1.5$~MHz~\cite{Rabi}. This indicates that the desirable
effective Rabi frequency
$\Omega=\Omega_0\eta^2=\delta_{\rm{ex}}\approx0.1$~MHz [for
implementing the operation in Eq.~(11)] could be experimentally
realized by applying the laser beam with proper power, as
$\Omega_0\propto\sqrt{P}$.

With the laser-induced switchable coupling we now discuss how to
implement the universal quantum gates by non-adiabatically
manipulations. For generality, we consider here $n\,(\geq2)$
potential wells (the adjacent wells are separated by an uniform
distance $d$), as shown in Fig.~3. Each potential well confines two
ions, e.g., $Q_n$ and $A_n$. The ion $Q_n$ (e.g., a $^{9}\rm
{Be}^+$~\cite{experiment1}) is used to encode the qubits by its
long-lived internal states $|\downarrow\rangle$ and
$|\uparrow\rangle$. While, the auxiliary ion $A_n$ (e.g., a
$^{40}\rm {Ca}^+$~\cite{experiment2}) is used for two purposes: (i)
cooling the CM $|\psi_{\rm cm}\rangle_n$ of the two ions $Q_n$ and
$A_n$ confined in the same trap, and (ii) implementing the
switchable coupling $\hat{U}_{n,n+1}(gt)$ between the CMs of the
ions trapped in adjacent potential wells.

\begin{figure}[tbp]
\includegraphics[width=9.5cm]{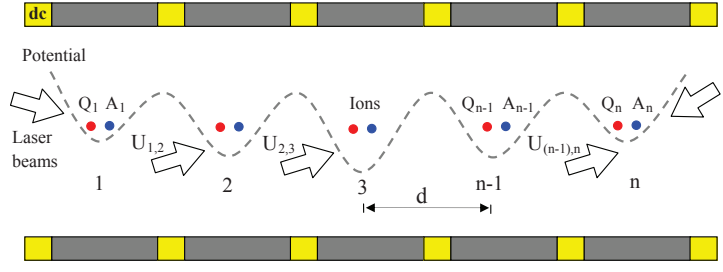}
\caption{(Color online) Sketch for implementing quantum logic
operations with $n\,(\geq2)$ potential wells. The distance between
the adjacent wells is $d$. Each potential well confines two ions:
the red ion (e.g. the ion $Q_n$) is used to encode the qubit, and
the blue ion (e.g., the ion $A_n$ ) is used to cool the ions'
center-of-mass mode (CM) and implement the switchable coupling
between the ionic vibrations in the adjacent potential wells.}
\end{figure}

In the same potential well the frequency of ions' breath mode (BM)
is significantly larger than that of the CM, so that we can treat
them separately~\cite{Steane,James}. Assuming the $n$ potential
wells have any different depths such that the CM-CM, BM-BM, and
BM-CM couplings between the different wells are naturally cut off
unless the laser-assisted coupling $\hat{U}_{n,n+1}(gt)$ is applied.
By selecting the proper sideband laser pulses the quantum
information (QI) between the internal states and CM can be
transferred, while the BM does not work (since it has different
frequency respect to the CM)~\cite{CZCNOT}. Therefore, one can use
only CM as the desirable data bus to realize the QI transformation
between the qubit trapped in the different potential wells.
Since we are considering two different ions (i.e., $^{9}\rm {Be}^+$
and $^{40}\rm {Ca}^+$) in each potential well, the laser
individually driving is unnecessary. The laser is required only to
be focused onto a waist smaller than $d\sim40$~$\mu$m (the distance
between the adjacent potential wells~\cite{experiment1}), in
principle.

It has been well-known that the single-qubit operations can be
exactly implemented by driving various resonant Raman
transitions~\cite{experimentgate,experimentstate}. The present
purpose is to implement the two-qubit controlled-{\small NOT} (CNOT)
gate~\cite{Cirac}
\begin{equation}
\hat{C}_{1,n}=\left(
  \begin{array}{cccc}
    1 & 0 & 0 & 0\\
    0 & 1 & 0 & 0\\
    0 & 0 & 0 & 1\\
    0 & 0 & 1 & 0\\
  \end{array}
\right)
\end{equation}
between the qubits (e.g., the ions $Q_1$ and $Q_n$) trapped in the
different potential wells. For simplicity, we suppose that the CMs
are initially prepared in the vacuum states $\prod_{j=1}^n|0_{\rm
cm}\rangle_j$. Then, the desirable CNOT gate $\hat{C}_{1,n}$ could
be realized by the following operational sequence:
\begin{equation}
\hat{C}_{1,n}=\hat{V}_1\hat{M}_{n,1}\hat{S}_{n}\hat{M}_{1,n}\hat{V}_1.
\end{equation}
Here, $\hat{V}_1$ represents an operation $|0_{\rm
cm},\uparrow\rangle_1\longrightarrow \exp[i(1-n)\pi/2]|1_{\rm
cm},\downarrow\rangle_1$ applied on ion $Q_1$ for transferring the
QI from its internal states to the CM in potential well $1$.
Experimentally, the operation $\hat{V}_1$ can be realized by
applying a red-sideband laser pulse to the ion, see, e.g.,
Refs.~\cite{CZCNOT,experimentgate}. The operation
$\hat{M}_{1,n}=\prod_{j=1}^{n-1}\hat{U}_{(n-j),(n-j+1)}(\pi/2)$
represents $n-1$ resonant pulses applied sequentially on the
auxiliary ions $A_2$, $A_3$, $\cdots$, and $A_n$ to couple the CMs
in the separated potential wells one by one. Physically, the
operation $\hat{M}_{1,n}$ transfers the QI stored in the CM of
potential well $1$ to that in the distant well $n$. Similarly,
$\hat{M}_{n,1}=\prod_{j=1}^{n-1}\hat{U}_{j,(j+1)}(\pi/2)$ transfers
the QI from the potential well $n$ back to the well $1$. Finally,
$\hat{S}_n$, a single-ion CNOT gate between the CM and internal
states of the ion $Q_n$, can be implemented by applying three
sequential laser pulses to the ion (as the demonstrations in the
experiment~\cite{experimentgate}).

The proposed CNOT gate $\hat{C}_{1,n}$ could be implemented within a
short operational time. First, the durations of operation
$\hat{V}_1$ and gate $\hat{S}_{n}$ by using the Raman schemes of
experiment~\cite{experimentgate} are estimated as $t_{\rm
v}\approx8\,\mu$s and $t_{\rm s}\approx50\,\mu$s, respectively.
Second, for the typical parameters~\cite{experiment1,experiment2}:
$d\approx40$~$\mu$m (the distance between the adjacent potential
wells) and $\nu_j\approx5$~MHz (the vibrational frequencies of the
CMs), the coupling strength between the adjacent CMs is calculated
as $g_{\rm cm}\approx35$~kHz, and thus the duration of the operation
$\hat{U}_{j,(j+1)}(\pi/2)$ is $t_{\rm u}\approx45\,\mu$s. Obviously,
all these durations (i.e., $t_{\rm v}$, $t_{\rm s}$, and $t_{\rm
u}$) are significantly shorter than the coherent times (which is
about $1$~ms) of the CMs~\cite{experiment1,experiment2,CZCNOT}.

Note that the laser coolings of CMs performed by using the auxiliary
ions do not affect the qubits. Therefore, during the operation
$V_1$, $\hat{S}_{n}$, or $\hat{U}_{j,(j+1)}$ applied on the ions
$1$, $n$, or $j$ and $j+1$, the other ions can be re-cooled. This
implies that the practical limit for the implementation of CNOT gate
$\hat{C}_{1,n}$ is the decoherence from the states
$|\downarrow\rangle$ and $|\uparrow\rangle$, i.e., the qubit.
Fortunately, the coherent times between the selected internal states
$|\downarrow\rangle$ and $|\uparrow\rangle$ are relative long, e.g,
up to $\sim10$~s~\cite{10s}. This allows us to implement the
desirable gates between a large number of trapped ions, e.g., $n\sim
20$. Indeed, the total duration $t_{\rm {total}}=2[t_{\rm
v}+(n-1)t_{\rm u}]+t_{\rm s}\approx1.8$~ms for generating the CNOT
gate $\hat{C}_{1,20}$ is far shorter than the coherent time of the
states $|\downarrow\rangle$ and $|\uparrow\rangle$.
Consequently, entangling a large number of ions (by using a series
of single-qubit operations and CNOT gates) may be possible.

In conclusion, we have proposed an alternative approach to implement
the universal quantum gates between the separately trapped ions.
Instead of the previous adiabatic operations of potential wells,
here the laser beams are utilized to non-adiabatically manipulate
the ion-ion couplings. Consequently, the desirable quantum gates
between the ions could be conveniently implemented by using only a
series of laser pulses. A key feature of our proposal is that it can
be generalized to the scaling ion-traps. With the experimental
parameters, e.g., the quadrupole transition Rabi frequency, the
vibrational frequencies of the trapped ions, and the distance
between the potential wells, we have analyzed the feasibility of the
approach in detail.

\vspace*{12pt}

\vspace*{12pt}

{\bf Acknowledgements}: This work was partly supported by the
National Natural Science Foundation of China Grants No. 11147116,
10874142 and No. 90921010, the Major State Basic Research
Development Program of China Grant No. 2010CB923104 (through Program
No. 973), and the open project of State Key Laboratory of Functional
Materials for Informatics.
\\
\\


\vspace{-1cm}
\begin{thebibliography}{99}
%
\bibitem{QC}
Ladd T D et al 2010 Nature {\bf 464} 45 
%
\bibitem{Cirac}Cirac J I and Zoller P 1995 Phys. Rev. Lett.
{\bf 74} 4091
%
\bibitem{warmION}M${\rm \o}$lmer K and S${\rm {\o}}$rensen A 1999 Phys. Rev.
Lett. {\bf 82} 1835
%
\bibitem{Physreports}
H$\ddot{\rm a}$ffner H, Roos C F and Blatt R 2008 Phys. Rep. {\bf
469} 155 
%
\bibitem{fourteenION}Monz T et al 2011 Phys. Rev. Lett. {\bf 106} 130506
%
\bibitem{scalable1}Kielpinski D, Monroe C and
Wineland D J 2002 Nature {\bf 417} 709 
%
\bibitem{scalable2}Home J P  et al 2009 Science {\bf 325} 1227 
%
\bibitem{Cirac2000}Cirac J I and Zoller P 2000 Nature {\bf 404} 579
%
\bibitem{experiment1}Brown K R et al 2011 Nature {\bf 471} 196 
%
\bibitem{experiment2}Harlander M et al 2011 Nature {\bf 471}
200 
%
\bibitem{Miao}Zhang M and Wei L F 2011 Phys. Rev.
A {\bf 83} 064301
%
\bibitem{IONSD}
Leibfried D et al 2003 Rev. Mod. Phys. {\bf 75} 281

%
\bibitem{SC}Bermudez A, Sch$\ddot{\rm a}$tz T and Porras D 2011 Phys. Rev. Lett.
{\bf 107} 150501
%
\bibitem{quadrupole}Roos C et al 1999 Phys. Rev.
Lett. {\bf 83} 4713
%
\bibitem{quadrupoleCPL}Liu Q et al 2011 Chin.Phys.
Lett. {\bf 28} 013201
%
\bibitem{CPL729}Shu H L et al 2007 Chin.Phys.
Lett. {\bf 24} 1217
%
%
\bibitem{Rabi}
Poschinge U G et al 2009 J. Phys. B {\bf 42} 154013 
%
\bibitem{Steane}Steane A 1997 Appl. Phys. B. {\bf 64}
623 
%
\bibitem{James}James D F V 1998 Appl. Phys. B. {\bf 66} 181 
%
\bibitem{CZCNOT}Schmidt-Kaler F et al 2003 Nature {\bf 422}
408 
%
\bibitem{experimentgate}Monroe C et al 1995 Phys. Rev.
Lett. {\bf 75} 4714
%
\bibitem{experimentstate}Meekhof D M et al 1996 Phys. Rev.
Lett. {\bf 76} 1796
%
\bibitem{10s}Langer C et al 2005 Phys. Rev. Lett. {\bf 95} 060502
%

\end{thebibliography}
\end{document}